# Developing a Dynamic Mobility Model for Backcasting Applications: A Case Study with Shared Autonomous Vehicles

*T. Héraud [1], V. Lakshmanan[1], A. Sciarretta[1]*

[1]*IFP Energies nouvelles, 1 et 4 avenue de Bois-Préau, 92852 Rueil-Malmaison, France (theotime.heraud@ifpen.fr, vinith-kumar.lakshmanan@ifpen.fr, antonio.sciarretta@ifpen.fr)*

# 1. Introduction

The European Union (EU) aims to achieve carbon neutrality by 2050, requiring a 90% reduction in transport emissions compared to 1990 levels [1]. As transport accounts for over a quarter of Europe's greenhouse gas emissions [2], effective governance, policies, and incentives are crucial for guiding the adoption of low-carbon technologies and modes of transport.

To identify optimal policy pathways, existing mobility models typically rely on forecasting approaches that simulate selected policy scenarios to assess their effectiveness in reducing GHG emissions. However, this reliance on predefined scenarios constrains the exploration of alternative outcomes, potentially preventing the identification of the optimal policy trajectory over time.

Backcasting is an optimisation-based approach that starts from a desired objective, such as emission reduction, and determines the optimal policy mix under given constraints. For a chosen set of levers, it guarantees identification of the optimal policy path over time. Initially applied qualitatively ([3]; [4]; [5]) or through static optimisation ([6]; [7]), it was later formulated as an optimal control problem by [8]. Their model optimised monetary incentives for electric vehicle adoption to minimise state budget while meeting $CO_2$ reduction targets, but focused only on passenger car fleet dynamics. This study extends the framework to a more complex case: introducing a shared autonomous vehicle (SAV) fleet within an urban mobility system.

Autonomous vehicles (AVs) are increasingly seen as inevitable, but their impacts on transport and GHG emissions remain uncertain. [9] shows that promoting electric SAVs could achieve greater decarbonisation than a carbon tax. AVs may also reshape modal choice by improving network fluidity and reliability [10]. They could further enhance safety, reduce costs, and increase efficiency; however, these benefits may be offset by rebound effects such as urban sprawl and higher demand for private AVs ( [11]; [12]; [13]). Yet, the literature lacks systematic quantification of these direct and indirect impacts, while forecasting remains uncertain and assumption-dependent, complicating policy design. The backcasting approach is particularly valuable for designing policy roadmaps to guide the adoption of this transformative technology in a desired direction.

This study models the introduction of an SAV fleet through a dynamic framework that supports both forecasting and backcasting, and proposes a method to identify desirable and undesirable policy effects within complex mobility systems. The model comprises six sub-models—mode choice, networks, infrastructure, level of service, vehicle stocks, and impacts—while transport demand is treated as exogenous. A nested logit model allocates demand between metro and car, and within cars between HVs and SAVs. Traffic flows are derived from a linear assignment model and fed back into mode choice, while fleet size determines emissions and operator costs. The model is tested on the Sioux Falls network [10] using forecasting simulations to analyse feedback mechanisms, followed by preliminary backcasting results outlining a policy roadmap to cut GHG emissions while minimising operator costs.

# 2. Modelling

The model developed is intended to represent the mobility system of a standard city with two types of network: a rail network (denoted by the subscript R in the variables) and a road network (subscript A). The only mode of transport on the rail network is the subway. On the road network, it is possible to travel with conventional vehicles, which we denote as Human-driven Vehicles (HV), or in Shared Autonomous Vehicle (SAV).

The model takes as inputs the overall transport demand defined by an Origin/Destination (O/D) matrix of flows, which is considered as exogenous, and the decision variable $u(t)$, which corresponds to the number of vehicles added to the SAV fleet at year $t$. The model outputs are the operator costs of SAV and rail, and the greenhouse gas (GHG) emissions at year $t$.

Various subsystems interconnect the model inputs and outputs. **Erreur ! Source du renvoi introuvable.** provides an overview of the model. It comprises six modules: (1) mode choice (by mode); (2) network (road and rail); (3) infrastructure (for road modes only); (4) level of service (SAV and rail); (5) vehicle stock (by mode); and (6) environmental and economic impacts. Also shown are a number of feedback loops, which will be analysed in detail in the next section.

In this section, we now detail the various sub-models that make up the mobility model.

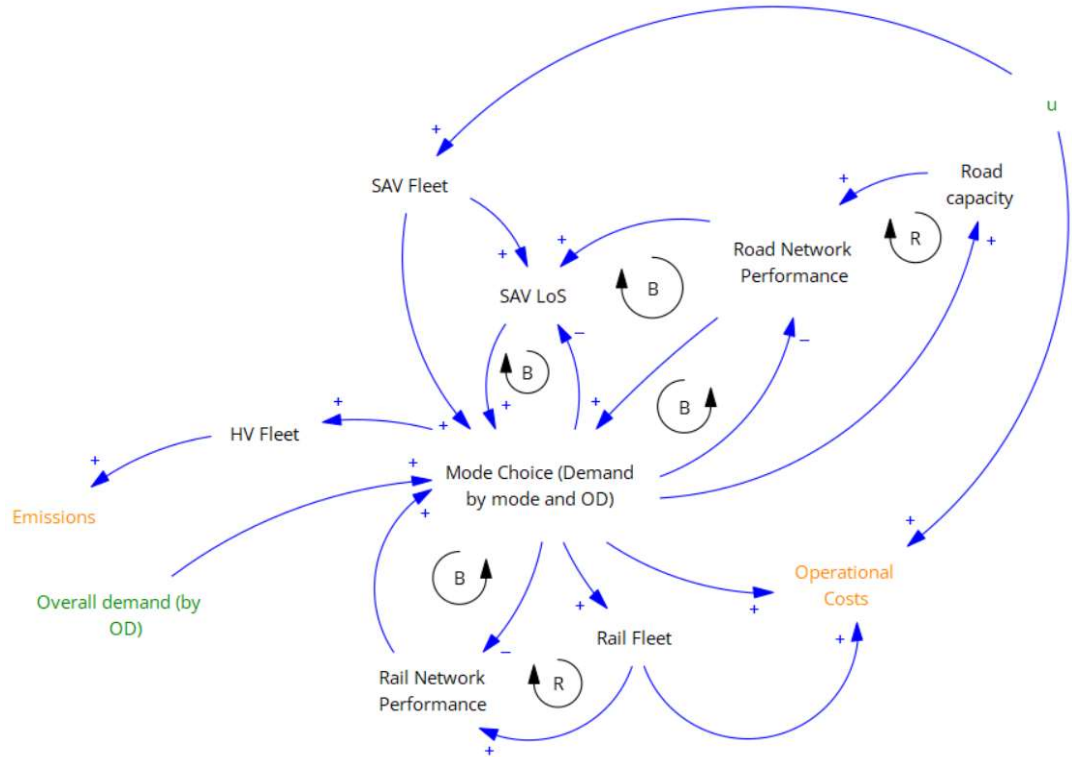

*Figure 1 - Causal loop diagram of the developed mobility model. Nodes represent model modules and the main variables associated with them; arrows represent functional connections between modules. Input variables are in green; outputs (impacts) in orange. Positive polarity of arrows denotes that an increase in the input node produces an increase in the output node, and vice versa. Feedback loops are explicitly represented and they can be balancing (B) when the overall loop polarity is negative or reinforcing (R) otherwise*

## 2.1. Mode choice

The "Mode Choice" sub-model evaluates the distribution of the overall demand by O/D, denoted $G$ (pax/h), among the different modes of transport. The input variables are the determinants of choices, i.e., the O/D travel times by car $t_A^{OD}$, those by rail $t_R^{OD}$, the average waiting time for SAV service $t_S^w$, and the access/egress time for rail service $t_R^{ae}$ depending on the rail frequency $F_R$. Other determinants such as the O/D distances are considered to be fixed parameters of the networks. The output variables are the demand levels (pax/h) for the three modes: $G_H$ (HV), $G_S$ (SAV) and $G_R$ (rail).

Mode shares are computed using a nested logit model, with HV and SAV grouped in the auto nest (A) and rail (R) separately. At the upper level, demand is split between nests A and R, then between HV and SAV within A. The model accommodates population segments with restricted mode access: (1) choice travellers (all three modes), (2) HV travellers (HV and SAV), (3) rail travellers (R and SAV), and (4) SAV travellers (SAV only).

The nested logit model uses mode-specific utility functions following the High-Tool formulation [14], combining time (in-vehicle and access/egress) and cost components (per km or fixed). Costs for HVs and rail are constant, while SAV costs vary with utilisation (see "Level of Service" sub-model) and are converted into time using the value of time (min/€). The utility parameters of the utility function are calibrated based on observed data for Ile-de-France region, see Sect.3.1.

## 2.2. Network

The network sub-model evaluates the in-vehicle travel time by mode and OD pair, $t_R^{OD}$ and $t_A^{OD}$ (min). The input variables are the demand by mode and OD pair (pax/h), and the network capacity per link $K_A$ (veh/h) and $K_R$ (pax/h). For both networks, the possible paths for each O/D pair are computed offline based on the solution of the User Equilibrium traffic assignment problem [15]. It is assumed that these paths do not change with a change in demand. Then, using results from [16] and the already known paths, the average O/D travel times can be evaluated as an affine function of O/D flows, based on the network topology and the link capacities. This calculation relies in particular on the commonly used BPR function, with distinct parameter values for road and rail networks taken from [13]. As mentioned above, these travel times are only in-vehicle travel times. The access/egress times are computed in the sub-model 'Level Of Service', see below.

## 2.3. Infrastructure

This sub-model computes the capacity of each link in the road networks.

SAVs are designed to exchange real-time traffic information with other SAVs and road infrastructure. However, according to [17], it can be assumed that connecting vehicles would increase network capacity by reducing the headways between vehicles. In a flow composed of HVs and SAVs, there are four possible combinations and four corresponding average headways: HV-SAV, HV-HV, SAV-HV and SAV-SAV. We assume that the average HV-SAV and HV-HV headways are equal, since in both cases a human driver would behave in the same way regardless of whether the vehicle in front is autonomous or not.

For this study, we adopt the formula described in [10] and [18], which allows the evaluation of the capacity of a link based on the fraction of SAV flow over total flow and the average headways.

This sub-model therefore takes the SAV and HV flow on each network link (obtained via the Network sub-model) as input, and outputs the associated capacity $K_A$ (veh/h).

## 2.4. Level of Service

The "Level of Service" sub-model evaluates for SAV and rail modes the access/egress times and cost, and the operating costs. These variables are used in the "Mode Choice" sub-model utility functions.

### 2.4.1. SAV

This sub-model takes the mean travel times by car, the SAV demand $G_S$, and the SAV stock $S_S$ as inputs. The output variables are the SAV waiting time $t_S^w$ (min), the customer operating cost $C_S^{c,op}$ (€/km), and the customer fixed access/egress cost $C_S^{c,ae}$ (€).

Waiting time is modelled with an M/M/s queuing system with finite population [19]. Customers (SAV requests) arrive according to a Poisson process, while service times (SAV rides) follow an exponential distribution. SAVs act as parallel servers, with the arrival rate depending on the number of customers already in the system and the service rate on the fleet size. Queuing theory then provides the probability distribution of customers in the system, the average queue length, and the effective arrival rate. With this information, an average waiting time depending on the SAV demand and the SAV fleet can be evaluated.

$C_S^{c,op}$ and $C_S^{c,ae}$ are both varying with SAV utilisation - the ratio of demand per SAV to a benchmark profitable mileage. Higher utilisation increases operator profitability, reducing customer costs. This relationship is modelled as a power-law (Cobb-Douglas) function with constant elasticity, as commonly used in transport economics to capture scale effects and capacity utilisation [20], [21].

To be included in the utility function of sub-model "Mode Choice", SAV waiting time and costs are passed through a first-order filter to introduce a lagging effect, thereby reflecting the customer's perceived waiting time and costs.

### 2.4.2. Rail

This sub-model calculates the access/egress times $t_R^{ae}$, i.e., the walking and waiting times.

The waiting time depends on the train frequency $F_R$ and a few given network parameters, such as the number of lines. The walking time is calculated based on the mean walking speed and the mean distance between two stations.

## 2.5. Vehicle Stocks

This sub-model computes the vehicle stocks for the different modes and the associated variables that are necessary to evaluate the impact in terms of GHG emissions and operator costs.

### 2.5.1. HV

The HV stock $S_H$ (veh) model is taken from [8], and only a brief description is given here. The stock is assumed to contain only two vehicle types distinguished by age classes, i.e, thermal and electric of ages 0 to 30. The inputs to the model include the annual vehicle-kilometres (vkm) travelled by HV. The vkm for HV is obtained by summing, over all O/D pairs, the product of the demand for HVs $G_H$, the corresponding O/D distance travelled, and the number of working hours in a year. Other exogenous inputs include the adoption coefficient (in a Bass adoption model for electric vehicles), the annual mileage $M$, the operating and purchasing costs of the two vehicle types, and the survival rate describing the natural obsolescence of the vehicle stocks. Each year, based on mileage, the current fleet, and the demand for vehicle-kilometres to be travelled, the number of new vehicles required is computed. Subsequently, a discrete choice model is used to define the proportion of thermal and electric vehicles.

### 2.5.2. SAV

The SAV stock $S_s$ (veh) is assumed to be composed of only electric vehicles. Its evolution is computed based on the surviving fleet at a given year and the number of new SAVs added to the stock. The latter is considered as input $u$ (veh/y), and the surviving fleet is computed based on a constant survival rate.

### 2.5.3. Rail

The rail stock model computes the rail stock $S_R$ (veh), the service frequency $F_R$ (veh/h) and the link capacity $K_R$ (pax/h). The input variable is the total rail demand $G_R$ (pax/h). The service frequency is assumed to be uniform across the network and is determined by dividing the total rail demand by the product of train capacity (pax) and the occupancy rate (dimensionless), both of which are considered known and constant.

The capacity of each link (pax/h) is supposed to be the same everywhere in the network and is equal to the product of the frequency of service and the train capacity.

The rail stock $S_R$ required to satisfy demand $G_R$ is determined by considering the mean length of a line, the average commercial speed of trains, and the desired service frequency $F_R$. The number of trains is then increased by a margin corresponding to the proportion of spare vehicles introduced as a reserve. This stock is used to calculate the operator costs, see below.

## 2.6. Impacts

This last sub-model computes the outputs of the whole mobility model: the GHG emissions in Mt/y, and the total operator cost in €.

### 2.6.1. Total operator cost

It is assumed that the SAV fleet and rail network are managed by a single operator, so the total cost combines both. SAV costs include operating and purchase components. The operating cost $C_S^{op}$ (€/vkm) covers energy and maintenance, scaled by total travel demand $D_S$ (vkm/y), and adjusted by a utilisation-dependent function with higher elasticity than for customer costs. Purchase cost corresponds to the unit vehicle cost $C_S^{pu}$ (€/veh) multiplied by the number of SAVs acquired annually,

$$C_S = C_S^{op} \cdot D_S + u \cdot C_S^{pu}$$

Rail costs consist of a fixed component $C_R^{fix}$ (€/y), covering network, station, technical systems, and staff, and a variable component proportional to travel demand $D_R$ (vkm/y). The latter includes operational costs $C_R^{op}$ (€/vkm) for energy, maintenance, and exploitation, as well as depreciation $C_R^{dep}$ (€/vkm) based on fleet size $S_R$ and a train purchase cost. The total rail cost then writes:

$$C_R = \left(C_R^{op} + C_R^{dep}\right) \cdot D_R + C_R^{fix}$$

### 2.6.2. GHG emissions

Tailpipe $CO_2$ emissions $E$ are represented using emission factors distinguished by age, $\epsilon_a$ (g/km) and the total annual mileage $M$ (km). We assume zero tailpipe emissions for electric vehicles. Consequently, the emissions contribution is solely from the thermal stock of HVs, distinguished by age classes $S_{H,T,a}$. The yearly emission $E(t)$ is computed as $E = \sum_a \epsilon_a \, S_{H,T,a} \, M$, and its cumulative emissions are represented by $\xi$.

# 3. Study Case

In this section, the model is calibrated and a case study to which it will be applied is selected. The aim will be to run forecasting simulations in order to identify the various loops involved in the model and to run initial backcasting simulations.

The model is applied to the Sioux Falls case from [10], using network data from [22]. The road network has 76 links serving all 24 nodes, with defined capacities, distances, and free-flow times; the rail network has 36 links across four lines serving 18 nodes. Nodes without rail are limited to HV and SAV. Transport demand (pax/h) is defined on 68 O/D pairs, assuming one passenger per vehicle on roads. Initial rail demand and capacities allow train frequency calculation, while the SAV fleet starts empty.

The utility functions parameters in the "Mode Choice" sub-model are calibrated to reproduce observed mode choice probabilities. Calibration uses the 2010 EGT survey for Île-de-France [23], in which all observed modes are aggregated into the three present model categories: rail (all rail modes), HVs (private and two-wheeled vehicles), and SAVs (treated as taxis), while active mobility and buses are excluded. HV and rail costs and average times/distances are derived from the High Tool model [14]. Only commuting trips are considered, with 66% of users assumed to own cars ("Choice Travellers") and 34% relying on rail (based on INSEE analysis [24]). Initial calibration showed very low SAV adoption because travel costs were initially set equal to conventional taxi fares [25]. To reflect the assumption that SAVs are more economical than taxis, SAV costs per km are then reduced by a factor of three and access/egress fees aligned with subway fares, resulting in consistent mode choice probabilities and increasing SAV adoption as fleet size grows.

Table 1 lists the main parameters of the above model and their values used for the simulation.

| Parameter | Value | Description |
|---|---|---|
| $x_i$ | [0.66,0,0.34,0] | Percentage of captive for each traveller type ([Choice traveller, HV traveller, Rail traveller, SAV traveller]) |
| $C_H^{c,ae}$ | 1.4 | (€), access/egress cost for HVs used in the utility function |
| $C_H^{c,op}$ | 0.0745 | (€/km), operating cost for HVs used in the utility function |
| $C_R^{c,ae}$ | 2.2 | (€), access/egress cost for rail used in the utility function |
| $C_R^{c,op}$ | 0.2263 | (€/km), operating cost for rail used in the utility function |
| $C_S^{c,ae}$ | 2 | (€), initial access/egress cost for SAVs used in the utility function |
| $C_S^{c,op}$ | 0.5 | (€/km), initial operating cost for SAVs used in the utility function |
| $C_R^{op}$ | 10 | (€/vkm), operating cost for rail operator |
| $C_R^{fix}$ | 5.4767e7 | (€/y), annual fixed costs for rail operator. |
| $C_S^{pu}$ | 120000 | (€/veh), SAV purchase cost |
| $C_S^{op}$ | 0.2 | (€/vkm), operating cost for SAV operator |

*Table 1 - Model parameters values used for simulation*

## 3.1. Forecasting results

The model is launched with a 15-year forecast. Overall demand is assumed to remain constant throughout the simulation period. Each year, 700 SAVs are added to the fleet. Figure 2 shows the evolution over time of the SAV fleet, demand for the three modes ("Mode Choice" sub-model), the evolution of road capacity linked to the use of SAVs and the evolution of train frequency linked to rail demand ("infrastructure" sub-model), as well as emissions and total annual cost (model outputs).

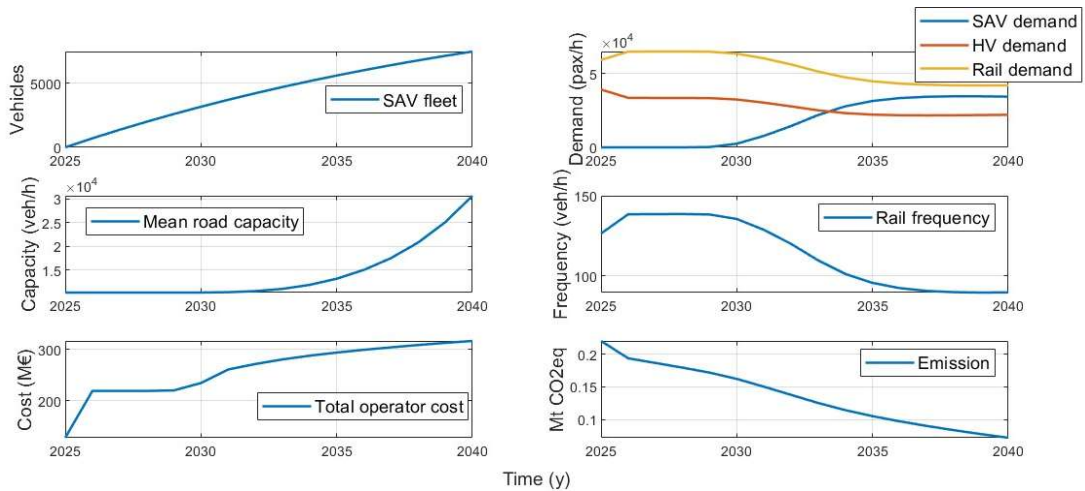

*Figure 2 - Forecasting results*

Figure 2 shows that increasing the SAV fleet improves service, boosting demand and road capacity, but reducing rail demand, frequency, and service quality—a potentially counterproductive effect given rail's carbon-free nature. SAV growth also lowers HV demand, and with HV fleet renewal, contributes to decarbonisation, accounting for ~8% of emission reduction compared to a "no-SAVs" scenario. Over the simulation, cumulative cost reaches 4.17 G€ and emissions 1.945 Mt. These dynamics illustrate the feedback loops involved and the challenge of designing an optimal policy $u(t)$ that balances emission reduction with minimal cost. The following section identifies these feedback loops to quantify their desirable and undesirable effects.

# 4. Analysis

In this section, we analyse the interactions between the different sub-models and propose a method for identifying the direct and indirect effects of the policy $u$.

## 4.1. Linear perturbation

Except for first-order dynamics in waiting time and SAV cost, system dynamics occur mainly at the input and output stages. Connections between SAV stock and HV demand form a system of static equations, whose solution defines a quasi-static equilibrium depending on SAV stock. Variations in the control $u$ propagate to other variables. Linearising the nonlinear equations around equilibrium yields 15 equations in 16 variables, with $u$ as the degree of freedom. Output dynamics, from HV demand to emissions, are also linearised locally. The causal loop diagram is complex and non-planar, with over 100 feedback loops and multiple feedforward paths, and is therefore not shown.

## 4.2. Simplified analysis of concurrent effects

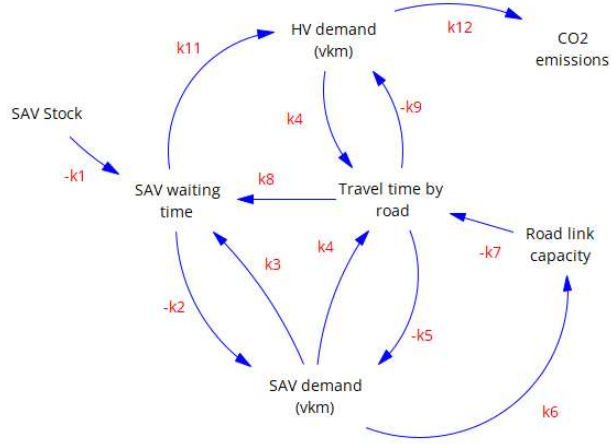

*Figure 3 – Simplified causal diagram*

In order to simplify the analysis, several simplifying assumptions are introduced. The variables $F_R$, $C_S^{c,ae}$, $C_S^{c,op}$ and $t_R^{OD}$ are supposed fixed. $G_R$ is given by the balance of demand, and first-order dynamics on waiting times and SAV costs are assumed to be absent. The system can then be represented as a causal graph where each relation between two variables is associated with a gain denoted $k_i$. Then, we obtain (see Figure 2) 8 feedback loops (3 reinforcing and 5 balancing), denoted $L_i$ and 3 feedforward paths (two decreasing, one increasing the emissions), denoted $P_i$.

Feedforward paths :

1. SAV stock – SAV waiting time – HV demand – Emission : $P_1 = -k_1 k_{11} k_{12}$ (negative)
2. SAV stock – SAV waiting time – SAV demand – Road travel time – HV demand – Emissions : $P_2 = -k_1 k_2 k_4 k_9 k_{12}$ (negative)
3. SAV stock – SAV waiting time – SAV demand – Road Capacity – Road travel time – HV demand – Emissions : $P_3 = k_1 k_2 k_6 k_7 k_9 k_{12}$ (positive)

Feedback loops :

1. SAV waiting time – SAV demand – SAV waiting time :   $L_1 = -k_2 k_3$ (balancing)
2. SAV waiting time –  SAV demand – Road travel time – SAV waiting time : $L_2 = -k_2 k_4 k_8$ (balancing)
3. SAV waiting time – SAV demand – Road capacity – Road travel time – SAV waiting time : $L_3 = k_2 k_6 k_7 k_8$  (reinforcing)
4. SAV waiting time – HV demand – Road travel time – SAV waiting time : $L_4 = k_{11} k_4 k_8$ (reinforcing)
5. SAV waiting time – HV demand – Road travel time – SAV demand – SAV waiting time: $L_5 = -k_{11} k_4 k_5 k_3$  (balancing)
6. SAV demand – Road travel time – SAV demand : $L_6 = -k_4 k_5$  (balancing)
7. SAV demand – Road capacity – Road travel time – SAV demand : $L_7 = k_6 k_7 k_5$  (reinforcing)
8. HV demand – Road travel time – HV demand : $L_8 = -k_9 k_4$  (balancing)

Such representations allow the use of Mason's gain formula [26], which enables quantifying the net effect of changes in SAV stock on emissions:

$$T = \frac{\Delta\ Emissions}{\Delta\ SAV\ stock} = \frac{P_1 - P_1 L_6 - P_1 L_7 + P_2 + P_3}{1 - \sum_{i=1}^{8} L_i + L_1 L_8}$$

The measure of the direct effect is the gain of the shortest feedforward paths, that is $P_1$ (negative). The indirect effects are measured by the other terms in the numerator, of which two are concurring with the direct effect ($P_2$ and $-P_1 L_6$), the others are directed in the opposite direction ($P_3$ and $-P_1 L_7$). In principle, there could be an undesired effect (introducing more SAV increases CO₂ emissions) if

$$P_1 - P_1 L_6 - P_1 L_7 + P_2 + P_3 > 0$$

that is,

$$-k_1 k_{12}(k_{11} + (k_4 - k_6 k_7)(k_{11} k_5 + k_2 k_9)) > 0$$

Therefore, a necessary but not sufficient condition is that $k_6 k_7 > k_4$.

This condition implies that the reduction in $t_A^{OD}$ from SAV-induced capacity increases would outweigh the rise caused by additional car demand—a situation unlikely to occur. Under the model's assumptions, the policy is therefore not expected to increase $CO_2$ emissions. Nevertheless, the presence of path $P_3$ and the reinforcing loop on HV demand $L_4$ may generate counterproductive effects. In addition, loops $L_3$, $L_4$ and $L_7$, which reinforce demand for HVs and SAVs, could further exacerbate this issue by diverting demand away from the carbon-free rail mode, as highlighted in the forecasting results.

# 5. Backcasting

This section illustrates the backcasting methodology to derive the optimal policy roadmap using the model described in Sect.2. The policy variable is the annual number of SAVs $u(t)$ introduced to achieve the total $CO_2$ emissions $\xi(T)$ less than the target $\bar{\xi}$, while minimizing the total operating costs of SAV and rail. The latter is given by

$$min \sum_t C_S(G_S, S_S, u) + C_R(S_R),$$

where $S_R$ is related to the state variable $S_{\hat{S}}$ through mode choice. With a 15-year time horizon starting in 2025, the reference scenario forecasts a total operating cost of 4.17 G€ and 1.945 Mt of $CO_2$ emitted, with a constant $u(t)$ of 700 SAVs per year. In the backcasting scenario, the total operating costs amount to 3.75 G€ representing a 10% cost saving while maintaining the same emissions as the reference scenario.

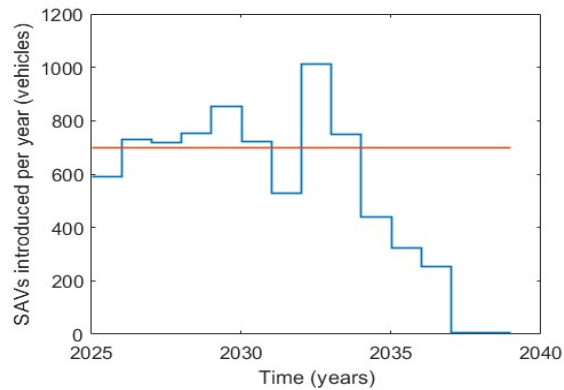

*Figure 4: Optimal SAV introduction schedule*

The optimal number of SAVs and its introduction schedule exhibit a variable behaviour as shown in Figure 4. The number of SAVs alternates around the reference value until 2031, after which a short dip occurs before 1,000 SAVs are introduced, followed by a gradual decrease until the end. This behaviour is considered optimal within the model assumptions. A full theoretical analysis of this behaviour is beyond the scope of this paper.

# 6. Conclusions

This study applies the backcasting methodology to a complex mobility system with multiple feedback effects, demonstrating its effectiveness in identifying an optimal policy that would be difficult to obtain under conventional forecasting approaches. The use case is the introduction of an SAV fleet. After detailing the sub-models, we presented forecasting analyses under an arbitrary policy, highlighting its effects on internal variables and identifying both direct and indirect impacts, including conditions under which undesirable effects arise. Preliminary backcasting results show a 10% reduction in operator costs compared to the reference forecasting policy for the same emission level. Over a 15-year horizon, SAVs yield an 8% emission reduction at an additional cost of 74%, corresponding to about 9,700 € per tonne of $CO_2$ saved—an expensive and seemingly ineffective measure. These results are indicative only, as they rest on strong assumptions and omit several parameters, including life-cycle impacts, charging infrastructure, and SAV-specific facilities. Future work will refine the analysis of feedback loops and backcasting outcomes.